\documentstyle[aaspp4,epsf,astrobib,11pt]{article}

\input{epsf}

\let\oldfootsep=\footnotesep
\def\lsim{\hbox{ \rlap{\raise 0.425ex\hbox{$<$}}\lower 0.65ex\hbox{$\sim$} }}
\def\gsim{\hbox{ \rlap{\raise 0.425ex\hbox{$>$}}\lower 0.65ex\hbox{$\sim$} }}

\def\msun { \rm {{\em M}_\odot}}

\def\umin{u_{\rm min}}

\def\tstar{t_{\rm *}}

\def\umin{u_{\rm min}}
\def\t0{t_{\rm 0}}

\def\vhat{\widehat{v}}

\def\kms {\,{\rm km \, s^{-1} }}
\def\kpc {\, {\rm kpc}}
\def\kmsk {\,{\rm km \, s^{-1} \kpc^{-1}}}
\def\spose#1{\hbox to 0pt{#1\hss}}
\def\simlt{\mathrel{\spose{\lower 3pt\hbox{$\mathchar"218$}}
     \raise 2.0pt\hbox{$\mathchar"13C$}}}
\def\simgt{\mathrel{\spose{\lower 3pt\hbox{$\mathchar"218$}}
     \raise 2.0pt\hbox{$\mathchar"13E$}}}

\begin{document}

\title{Observations of the Binary Microlens Event MACHO-98-SMC-1
          by the Microlensing Planet Search Collaboration}
\author{
  S.H.~Rhie\altaffilmark{1},             
  A.C.~Becker\altaffilmark{2,3},         
  D.P.~Bennett\altaffilmark{1,3},      
  P.C.~Fragile\altaffilmark{1},           
  B.R.~Johnson\altaffilmark{5}            
  L.J.~King\altaffilmark{1,3},           
  B.A.~Peterson\altaffilmark{4},         
  J.Quinn\altaffilmark{1}           
\begin{center}
{\bf (The Microlensing Planet Search Collaboration) }\\
\end{center}
}

\altaffiltext{1}{Department of Physics, University of Notre Dame, Notre Dame, IN 46556
}
 
\altaffiltext{2}{Departments of Astronomy and Physics,
  University of Washington, Seattle, WA 98195
}

\altaffiltext{3}{Center for Particle Astrophysics,
  University of California, Berkeley, CA 94720
}

\altaffiltext{4}{Mt.~Stromlo and Siding Spring Observatories,
  Australian National University, Weston, ACT 2611, Australia
}

\altaffiltext{5}{Tate Laboratory of Physics, University of Minnesota, 
  Minneapolis, MN 55455}
  
\setlength{\footnotesep}{\oldfootsep}
\renewcommand{\topfraction}{1.0}
\renewcommand{\bottomfraction}{1.0}     


\vspace{-5mm}
\begin{abstract} 
\rightskip = 0.0in plus 1em

We  present the observations of the  binary lensing event
MACHO-98-SMC-1  conducted at the Mt.~Stromlo 74" telescope
by the Microlensing Planet Search (MPS) collaboration. 
The MPS data constrain the first caustic crossing to have occurred 
after 1998 June 5.55 UT and thus directly rule out one of the two
fits presented by the PLANET collaboration (model II).
This substantially reduces the uncertainty in the the relative proper 
motion estimations of the lens object.

We perform joint binary microlensing fits of the MPS data together with 
the publicly available data from the EROS, MACHO/GMAN and OGLE 
collaborations.   We also study the binary lens fit parameters
previously published by the PLANET and MACHO/GMAN collaborations by using 
them as initial values for $\chi^2$ minimization.  Fits based on
the PLANET model I appear to be in conflict with the 
GMAN-CTIO data.  From our best fit, we find that the lens system has a 
proper motion of $\mu = 1.3\pm 0.2 \kmsk$ with respect to the source, 
which implies that the lens system is most likely to be located in the 
Small Magellanic Cloud strengthening the conclusion of previous reports. 

\end{abstract}
\vspace{-5mm}

\newpage
\section{Introduction}
\label{sec-intro}

The Microlensing Planet Search (MPS) Project monitors microlensing
events discovered in progress by the EROS, MACHO, and OGLE experiments
in search for the microlensing signature of planets orbiting faint lens 
stars or ``non-standard" microlensing  light curves which
can provide an additional constraint on the distance and mass
of the ``dark" lens systems. The MPS project primarily
monitors lensing events toward the central regions of the Galaxy where
the microlensing events are most numerous. However,
``non-standard" events detected towards the Magellanic Clouds present a unique
opportunity to learn about the composition of the dark halo that 
dominates the mass of the Milky Way, and these events are observed at a high 
priority.  The binary microlensing event MACHO-98-SMC-1 was one such case.

The measurements of the microlensing optical depth towards the Large Magellanic
Cloud (LMC)  indicates that there is a previously unknown
``dark lens population" toward the LMC  \cite{macho-lmc2}. 
If the microlensing population is dominated by Galactic halo objects, 
the time scale of the microlensing events indicates their typical mass to be 
$\sim 0.5 \msun$, which may be low mass stars, white dwarfs, or primordial
black holes \cite{bhole}.   A large population of low mass stars or white 
dwarfs in the Galactic halo would likely have other observable effects, 
and it has been speculated that the LMC microlensing events are due to
normal stars in the LMC itself \cite{sahu}.  
The possible confusion between LMC self-lensing and lensing by Galactic 
halo objects derives from the fact that  the distance
and the mass of the lensing objects cannot be directly measured for most of the 
microlensing events.  For a ``standard" microlensing event, the only 
constraint on the three unknowns of distance, velocity and mass of the lens 
system comes from a single observed quantity, 
the ``Einstein ring radius crossing time" $t_E$.  

In a caustic crossing binary lensing event, 
one can measure one more independent
parameter, namely, the ``source radius crossing time", $t_{\ast}$,  and thereby 
estimate the relative proper motion $\mu$ of the lensing object with respect 
to the source star by independently determining the angular size of the  source 
star from its brightness and color.
A measurement of the relative proper motion, $\mu$, allows the determination of
the angular Einstein ring radius, $\theta_E = \mu t_E$.   Once $\theta_E$  is
known,  the mass of the lensing object is expressed as 
a simple monotonic function 
of the distance to the lens (if the distance to the source is known). 
If $D_\ell$ and $D_s$ are the distances to
the lens and the source star, and $\delta \equiv D_\ell/D_s$,  then 
\begin{equation}
 \left({M\over M_{\odot}}\right) =  {\delta \over 1 - \delta}\,
   \left({D_s\over 60\kpc}\right) 
      \left({\theta_E\over 0.369\,{\rm mas}}\right)^2 \ . 
  \label{eq-M}
\end{equation}
$D_\ell$ is not known, but it is strongly correlated with the 
proper motion, $\mu$.  For example,
if we take our best fit value of $t_E = 70.5\,$days and assume
$D_s = 60 \kpc$ and $\mu = 1 \kmsk$, then the lensing object will be 
a binary in the SMC with the total mass $M \approx 0.36 M_{\odot}$ 
for $D_s-D_\ell = 2 \kpc$.   
For a typical halo lens we expect $D_\ell \approx 10 \kpc$ and
a transverse velocity of $\approx 200 \kms$ 
assuming a standard isthermal sphere halo model \cite{gdynamics}.
This yields $\mu \approx 20 \kmsk$ for a typical halo lens (which would imply
a lens mass of $M = 0.81 M_{\odot}$ from eq.~\ref{eq-M}.)
Of course, in order to compare to our measurement of $\mu$, we should
compare to the predicted $\mu$ distributions for halo and SMC lenses.
This has been done for some simple SMC and halo models by
\citeNP{smc-nbody,98smc1-planet,98smc1-macho,honma}, and their results
indicate that for most values of $\mu$, either a halo or SMC lens is
strongly preferred. However, depending on the halo and SMC models used,
there is an overlap region at $\mu = 2-4 \kmsk$ which is marginally
consistent with either a halo or SMC lens at the $2-3\sigma$ confidence level.
(\citeN{honma} also points out a selection effect that will tend to bias
$\mu$ measurements towards smaller values.)
In the case of MACHO-98-SMC-1, model II of the PLANET collaboration
\cite{98smc1-planet} yields $\mu = 2 \kmsk$ which does not allow 
a definite determination of the lens location in the
in the halo or in the SMC \cite{honma}.    

\begin{figure}
\plotone{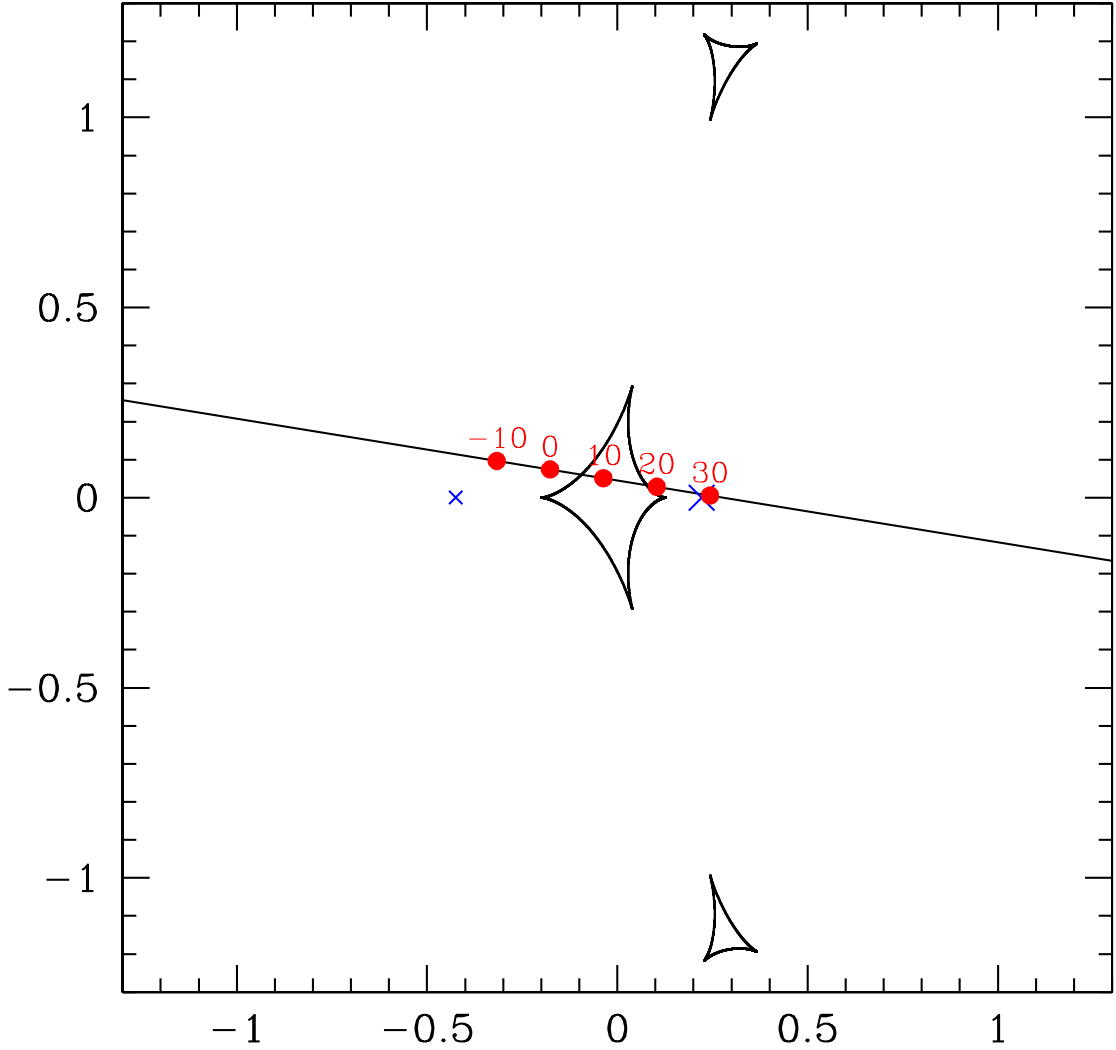}
\caption{This figure shows the configuration of the caustic curves for the
MPS lightcurve fit to binary lensing event MACHO-98-SMC-1. The crosses
indicate the locations of the lenses, and the straight line indicates the
path of the source star with respect to the caustic curves. The red dots on
the source star path indicate the location of the source at various dates
given in June, UT. The distance scale for the axes is the Einstein ring
radius, $R_E$. Note that the actual size of the source star is only about
$0.0015R_E$ that is much less than the thickness of the curves in the
Figure.
  \label{fig-caustic} }
\end{figure}

The main features of a binary lensing event are determined by the location
of the caustic curve in the source plane indicates the location of the
source with respect to the lens system projected to the position of the source.
The caustic curve is where the  number of images of the source changes by
two.  In binary lensing,  the caustic curve is made of one, two, or three
closed curves,  and the number of images is 5 inside the closed curves and
3 outside.  The caustic curves for MACHO-98-SMC-1 (according to the MPS fit) 
are shown in Figure \ref{fig-caustic}.  
When the source moves inside one of these
caustic curves, two new images are created, and the magnification of these new
images is singular at the point of the caustic crossing.  Because of this
discontinuity (intrinsic width zero),   the finite angular
size of the source star is {\it necessarily} resolved during a caustic 
crossing.  At the same time, this discontinuity makes it difficult to observe 
the first caustic crossing (going into the caustic).  However, there is always
the second opportunity to monitor a caustic crossing once the first caustic
crossing has occurred because of the closedness of the caustic curve, and   
the second caustic crossing (exit from the caustic) time can be predicted
through real-time data reduction and binary lens fitting as 
the source proceeds inside the caustic. The timely pre-caustic crossing 
announcements from the MACHO/GMAN group \cite{macho-iauc98a,macho-iauc98b}
allowed intense monitoring of the second caustic crossing of   
the MACHO-98-SMC-1  by the microlensing community around-the-clock from 
all three (temperate) continents of the Southern Hemisphere
\cite{98smc1-eros,98smc1-planet,98smc1-macho}. This resulted in a 
light curve which is well sampled in the second caustic crossing region.

According to our fit,  the
binary lensing event MACHO-98-SMC-1 was magnified by $\approx 70$ times at
the maximum of the second caustic crossing.  Such extreme magnification
is also useful in studying the properties of the lensed star
\cite{lennon,95blg30}.   In order to obtain an accurate model of the lensing
event, which is necessary to determine $\mu$, however,  it is not enough to 
have only meticulous measurements of the second caustic crossing.    
The main contribution of the MPS data is to constrain the time of the 
poorly sampled first caustic crossing and directly rule out the ``outlier"
PLANET model II.

\section{MPS Observations and a Constraint on the First Caustic Crossing}
\label{sec-obs}

The Microlensing Planet Search project has been allocated approximately
100 nights on the Mt.~Stromlo Observatory (MSO) 1.9m telescope for the 1997
and 1998 Galactic bulge seasons. 
Ongoing microlensing events announced by the
MACHO, OGLE, and EROS collaborations are monitored at intervals of 1-2 hours
using the Monash Camera which is a Cassegrain imager fitted with
a SITe 15 micron 2048 x 4096 AR-coated thinned CCD. The data is reduced 
within a few minutes after it is taken using automated Perl scripts written
by one of us (ACB) which call 
a version of the SoDOPHOT photometry routine \cite{sod}. This allows the 
immediate discovery of any unusual microlensing features that might be 
in progress.

MPS made its first observation of event MACHO-98-SMC-1 
about one day after MACHO microlensing alert issued  May 25.9 UT
and continued its observations as a medium priority target.   
One of these observations was obtained at June 5.549 UT which turned out to
be the last observation prior to the caustic crossing.
After the caustic-crossing binary lensing alert issued June 8.99 UT, 
MACHO-98-SMC-1 was upgraded to a high priority target.  
However, we were not scheduled on the MSO 1.9m until June 18, so 
our coverage of the event while 
the source was inside the caustic curve was minimal.  
On the 18th, the imager was available again, and the MSO staff kindly
altered the telescope pointing limits to allow us to observe the SMC
almost completely under the pole at an airmass of 3.2.
We made the first observation at 
June 18.332 UT about 40 minutes after the trailing limb of the star cleared
the caustic (according to our best fit which indicates the second caustic
crossing endpoint at June 18.304 UT). Although we 
missed the second caustic crossing, we kept MACHO-98-SMC-1 at a high priority 
to cover the ``cusp approach" lightcurve feature. This was a rise to
a gentle peak and subsequent decline that occur as the source passes in front
of one of the sharp ``cusps" of the caustic curve 
(see Figure \ref{fig-caustic}).
Good coverage of this feature is important if we hope to constrain the 
global parameters of the lensing event.
Unfortunately, due to poor (la Ni\~na) weather, our coverage of the
``cusp approach" is not very good.

The intense worldwide monitoring of the event was 
concentrated around the second
caustic crossing making it the best covered caustic crossing in microlensing 
history.  However, a reasonable amount of data around the first caustic crossing
is necessary to pin down the lens parameters.
The OGLE observation June 6.40 UT and the MACHO/GMAN observation 
at June 6.45 UT that indicate that the
first caustic crossing must have occurred by June 6.0 or so.
A lower limit on the time of the first caustic crossing is set by the MPS
observation at June 5.549 UT which is the last observation before 
the first caustic crossing. The measured flux of this MPS observation is 
consistent with the slow variation of the lightcurve for a source approaching 
a binary caustic prior to the first contact of the caustic with the
stellar limb. Thus, the first caustic crossing is 
constrained to have been completed  within the window 
of $\sim 20$ hours between June 5.55 - 6.40 UT. 

\section{Binary Lensing Analysis}
\label{sec-ana}

A binary lensing event involves seven parameters. These include three parameters
that also exist for single lens events: the Einstein ring crossing time,
$t_E$, the ``impact distance\rlap," $\umin$, from the origin of the coordinate 
system, and the time of the closest approach to the origin, $\t0$.  
We choose the lens system center of mass (c.m.) as the origin so that 
$\t0$  would be the most reasonable generalization 
of the time for the maximum amplification of a single lens.  (The c.m. resides 
inside the caustic here. It always does when $a \le \sqrt{2}$.) This would
also be the most convenient coordinate system if we were to consider
the lens system orbital motion.
There are three additional parameters intrinsic to a binary lens:
the fractional mass, $\epsilon$, of the first lens, the lens separation $a$,
and the intersection angle of the source trajectory with the lens 
axis, $\theta$. (The first lens is the one on the left in Figure
\ref{fig-caustic}). The final parameter is
the source radius crossing time $\tstar$ which is obviously critical for the
lens proper motion determination.

In addition to these microlensing parameters, we must have additional
parameters to describe the unlensed brightness of source star in each
pass band, from each observing site (since the instrumental pass bands from
different telescopes are never identical). Also, since the microlensing
events are found in crowded stellar fields, it is usually the case that
the lensed source is blended with other unlensed sources that happen to
fall within the same seeing disk. Thus, we require an additional parameter
for the brightness of any unlensed sources which are blended with the lensed
source.  These parameters need not be included for the non-linear $\chi^2$
minimization process, however, because the observed flux depends linearly
on  the   brightness of lensed star and its unresolved companions.  
Our $\chi^2$ calculation routine automatically 
minimizes $\chi^2$ with respect to these linear parameters for every
set of intrinsic microlensing parameters that is considered. This makes
our fitting routine converge to the best fit much more quickly than it would
if these were included as nonlinear fit parameters. However, it also
complicates the interpretation of our error estimates because the
error estimates for the blending parameters are calculated with the
intrinsic lensing parameters held fixed.

When a source is inside a caustic curve, there are two extra images in addition 
to the three ``normal" images, and when the caustic curve crosses the
source star, the two extra images are only partial images joined together
along the critical curve.  The time it takes for the stellar diameter to
cross the caustic, $2\Delta t$, can be measured using only observations
near the time of the caustic crossing.
However, $\tstar$ can be determined from $\Delta t$ only if we know 
the angle, $\phi$, between the source trajectory and the caustic curve at the 
crossing: $\tstar = \Delta t ~\sin\phi$. $\phi$ can only be determined
by a fit to the entire microlensing lightcurve, so measurements of
the caustic crossing alone are not sufficient to determine $\tstar$.
It is possible to constrain $\tstar$ without a determination of $\phi$
\cite{98smc1-eros}, but this constraint may be very weak.

The modeling of a binary lensing event presents a number of difficulties.
First, the caustic crossings mean that binary lensing lightcurves generically
have very sharp features, and since the photometric measurements discretely
sample the lightcurves, there can be large changes in $\chi^2$ caused by
small changes in the parameters that happen to move a caustic past
the location of a data point. The singular nature of microlensing
magnification also causes difficulties for the integrations necessary
to calculate the microlensing magnification of a finite size source star
and prevents the use of fast high order methods \cite{rhie-benn99}.

Yet another difficulty with binary lens fits is that the location of the
caustic crossing in the lightcurve depends in a complicated way on the
microlensing parameters. The time of the caustic crossings can generally
be pinned down to reasonable accuracy simply by inspection of the microlensing
lightcurves, but it is difficult to translate this into a constraint on the
microlensing parameters: $\epsilon$, $a$, $\theta$, $\umin$, $\t0$, and
$t_E$. However, since the times of the caustic crossings can readily be
calculated for any set of parameters, it is possible to shift $\t0$ and
rescale $t_E$ to put two caustic crossings at specified locations in time.
We use such a procedure to replace the parameters $\t0$ and $t_E$ by
the first and second caustic crossing times, $t_{cc1}$ and $t_{cc2}$, for
many of our binary lens fits.

The $\chi^2$ minimization for our microlensing fits is carried out with
the aid of the MINUIT routine \cite{minuit}. The fitting proceeds in several
stages. First, in order to find candidate global fits, we take the data
sets and remove many of the data points from regions where the data highly
oversample the lightcurve features in order to speed up the calculations
in the early phases of the fitting process. We also remove all of the 
data points which resolve the caustic crossing so that the search for
candidate global microlensing fit parameters can be done in the point
source limit which typically speeds up the calculations by a factor 
of 10 or more. We then start a number of Monte Carlo parameter searches
to find good starting points for the microlensing fits using MINUIT's
SEEK routine. During the Monte Carlo parameter searches, the values
of $t_{cc1}$ and $t_{cc2}$ are constrained to small time intervals
which were determined by inspection of the individual lightcurves.
This results in a number of candidate microlensing models which
are passed to the second stage of the fitting procedure.

In the second stage of the fitting process, we include some of the data
which resolves the caustic crossing and to fit all of the candidate 
microlensing models again with a finite value for $t_\ast$. This procedure
converges to the final fit much more quickly than if all the
data were used at this stage. Once the finite source effects are included,
it is necessary to take the limb darkening of the source into account. 
For our preliminary fits, we have used a standard ``linear" limb darkening
model, but we have also used the ``square-root" model advocated by 
\citeN{diaz-limbd}
at the stage of the final fits which use the full data set.
The limb darkening coefficients were taken from
\citeN{claret-limbd} and \citeN{diazetal-limbd}.

In addition to this procedure used to find new fits, we have also
tried fits using initial conditions based upon the fits reported by
the PLANET and MACHO/GMAN collaborations.

\subsection{Previous Observations, Analyses, and Fits}
\label{subsec-prev}

Observations of MACHO-98-SMC-1 have been previously presented by
the EROS, PLANET, MACHO/GMAN, and OGLE collaborations. 
\cite{98smc1-eros,98smc1-planet,98smc1-macho,98smc1-ogle}  The EROS
observations from La Silla  covered a significant fraction of the 
falling curve of the second caustic crossing through the caustic crossing
``end point" and several hours beyond, and it was the first time 
that the linearity towards the 
``end point" was  observed.   At the ``end point", the source star 
completely exits the caustic, and the additional two bright partial 
images vanish,  causing the curvature of the light curve to change 
abruptly.   The ``end point" was estimated to have occurred   June 18.297 UT. 
From the linearity spanning 1.8 hours, the EROS collaboration suggested
a constraint $\mu \sin\phi \lsim 1.5 \kmsk$. Since they reported on data
only from the night of the second caustic crossing, EROS was not able to
determine the caustic crossing angle $\phi$,
so their constraint on the lens proper motion was weak.
However, the EROS data has the best coverage of the
caustic crossing ``end point" which proves very valuable 
when combined with other data sets.

The PLANET collaboration monitored the event since shortly after the
binary lens alert and 
had excellent coverage of the second caustic crossing peak turn-over 
from the SAAO 1m.  They also measured the spectrum at the light curve
peak from the SAAO 1.9m. 
They presented two binary lens fits, which we will refer to as
PLANET-I and PLANET-II, that resulted in 
$\tstar = 0.122$ and $0.0896$ days. The models 
PLANET-I and II differ by $\sim 58$ in $\chi^2$ which is formally a
$7.6\sigma$ deviation. However, the $\chi^2$ per degree of freedom for
each were fairly large (2.37 and 2.73 respectively), and
they argued that both the fits should be considered to be 
viable fits (to account for unspecified systematic errors).
     
\begin{figure}
\plotone{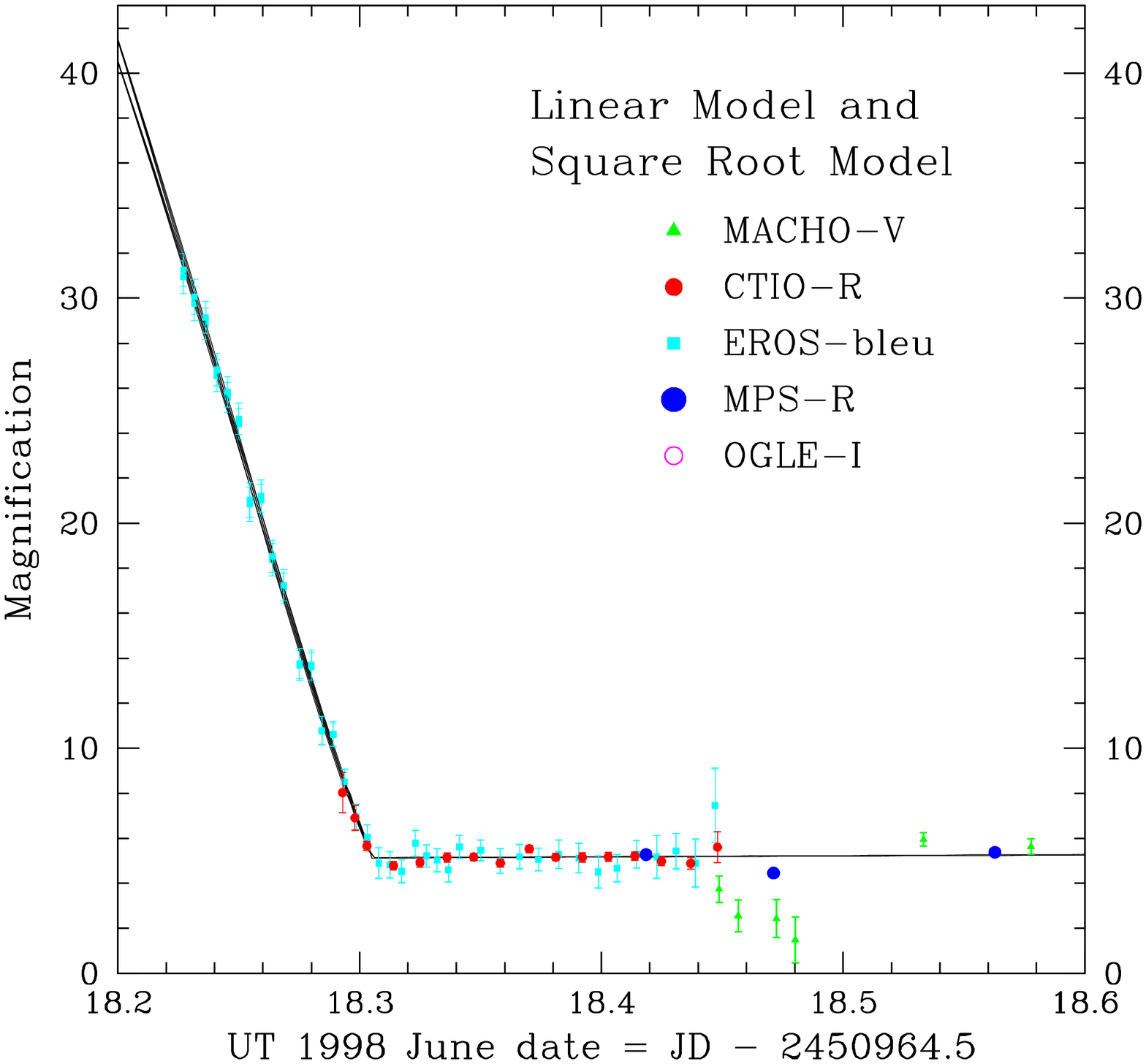}
\caption{This figure shows the 2nd caustic crossing endpoint MPS fit using 
the square root and linear limb darkening models. The square root model is
the one that predicts a slightly lower magnification at June 18.20.
  \label{fig-endpoint} }
\end{figure}

The MACHO/GMAN group reported their data from the Mt.~Stromlo
1.3m and the CTIO 0.9m telescopes \cite{98smc1-macho} and  presented a binary
microlens fit to the data combined with the EROS data.
Their fit differed from both PLANET-I and PLANET-II,
and MACHO/GMAN suggested that both the PLANET models 
might be inconsistent with pre-caustic-crossing MACHO/GMAN data.
Their estimate of the source radius crossing time
was $t_{\ast} = 0.116$ days.  The CTIO 0.9m observations registered the 
caustic crossing ``end point" 
at $\approx$ June 18.304 UT which agrees with the EROS data reduced 
with SoDOPHOT  (see figure~\ref{fig-endpoint}).  The MACHO/GMAN fit 
indicates that the second caustic crossing peak amplification 
was $\approx 70$ while PLANET-I indicates that the maximum 
amplification was $\approx 100$.  The main difference here is that the 
the PLANET-I indicates a fainter source star with more of the
baseline flux coming from unlensed stars.

The OGLE collaboration reported their data from Las Campanus 
(1.3m Warsaw telescope) that includes the first observation after the first 
caustic crossing at June 6.40 UT. They did not perform any microlensing fits, 
but they suggested that model PLANET-I is more consistent 
with the OGLE data than PLANET-II.  They also suggested that MACHO/GMAN fit
may be off by 0.14 days for the first caustic crossing.  In 
the MACHO/GMAN fit, 
the first caustic peak crossing occurred at $\approx$ June 6.24 UT, and hence, 
the suggestion by the OGLE team corresponds to the first caustic peak
crossing at  $\approx$ June 6.10 UT.  In model PLANET-I,  the peak crossing
time was $\approx$ June 6.08 UT, and thus, the OGLE team concluded that 
the OGLE data is probably most consistent with model PLANET-I.   

\subsection{MPS fits, Analyses, and Comparison}
\label{subsec-mps-fit}

In this section we present our binary microlensing fit results for
the data set including the MPS data plus the publicly available 
MACHO/GMAN, EROS, and OGLE data, and we interpret the meaning of these
results. We assume that the source star is a single lens star which was
lensed by a binary lens with no significant orbital motion.

\begin{figure}
\plotone{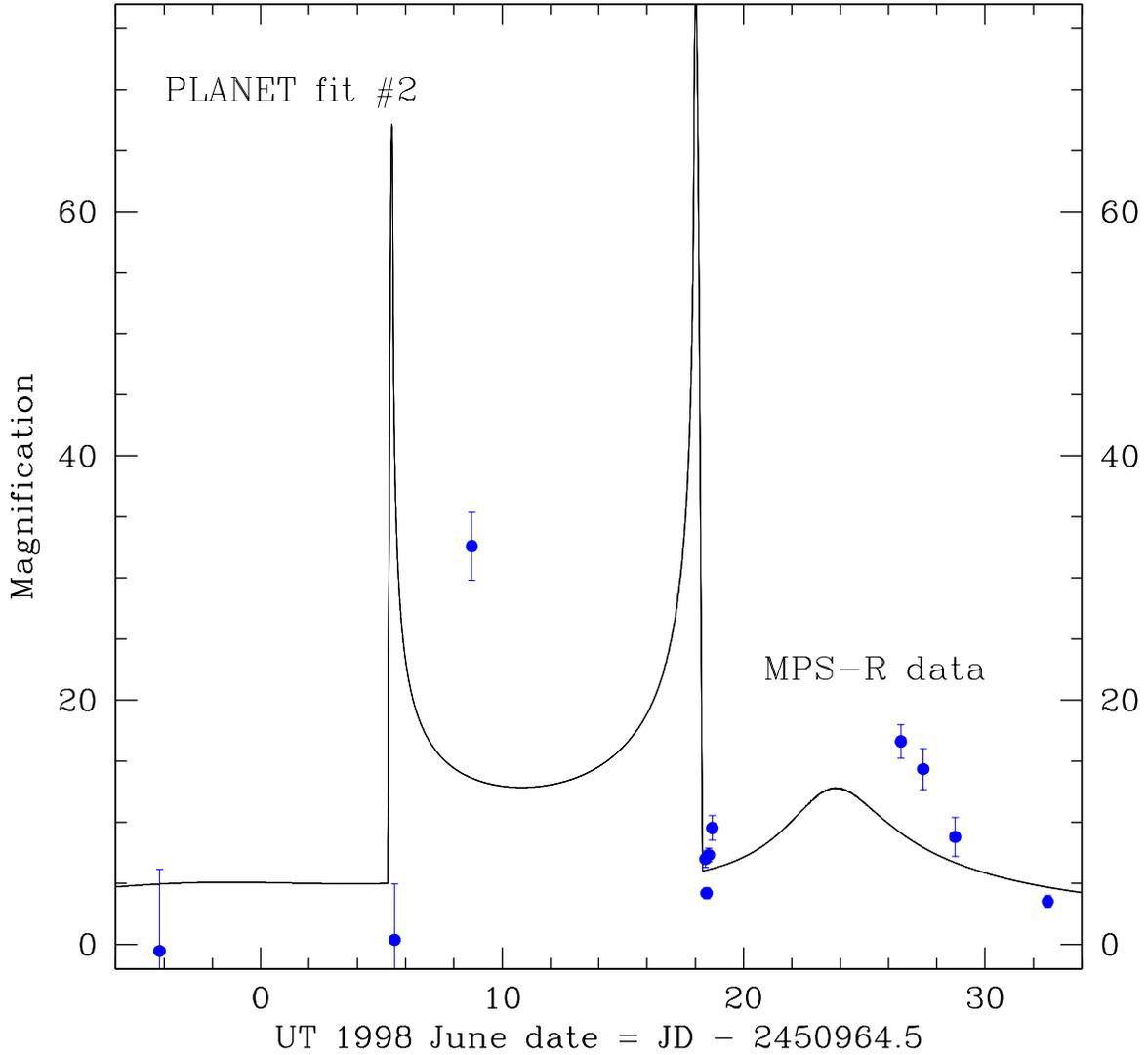}
\caption{This figure shows a comparison of the MPS data to the PLANET-II fit.
We have allowed the fluxes of the source star and any unlensed stars in
the same seeing disk to take the values which give the lowest $\chi^2$
value. The observation at June 5.55 UT indicates that the caustic crossing
had not yet begun, contrary  to the PLANET-II model prediction.
The attempt to fit this point results in a ``best-fit" curve which does not
agree with most of the other data points. 
  \label{fig-plan2} }
\end{figure}

The most obvious result of the MPS observations is that the PLANET-II
model is ruled out.
The MPS observation at June 5.55 UT indicates that the leading limb of the
source star has not yet crossed the caustic. This is inconsistent with 
the PLANET-II model which predicts the leading limb to cross the caustic at
June 5.25 UT, the stellar center ``caustic crossing time" at June 5.36 UT,
and the first caustic crossing lightcurve peak to occur at June 5.43 UT. 
Figure \ref{fig-plan2} shows a comparison of the PLANET-II fit to the
MPS data. In order to put it into the statistical perspective, 
we normalize the MPS data  to the PLANET-II fit using the
34 other observations (which do give an acceptable fit to the data), and 
the PLANET-II prediction for June 5.55 UT exceeds the observed brightness
by 29$\sigma$. Thus, the PLANET-II model is clearly ruled out.
Note that in Figure \ref{fig-plan2} and in all subsequent plots, 
the MPS data have been 
binned into nightly bins for all nights with multiple observations except
for the night of June 18 where 16 observations have been grouped into 4 bins.

The MPS observation on June 5.55 along with the OGLE
observation at June 6.40 UT and the GMAN-CTIO observation at June 6.45
constrain the caustic crossing to have occurred close to June 6.0 UT.
The MPS fit to the combined data set provides an acceptable fit to the
data near the first caustic crossing and indicates 
that the  first ``caustic crossing time" was June 5.91 UT,
and PLANET-I and MACHO/GMAN also seem consistent with this data
within the limit of the poor 
coverage.  Therefore, we will focus on a comparison 
between the MPS, MACHO/GMAN and PLANET-I fits as well as the lightcurve
details of the second caustic 
crossing where we hope to reconstruct the ``missing peak\rlap." (A future
comparison with the PLANET data should test our ability to predict the
features of the second caustic crossing peak from the other data sets which
do not sample the peak.)

Tables~\ref{tab-lpar}-\ref{tab-cpar} shows the summary of the results of the
microlensing fits we have performed on the combined EROS/GMAN/MACHO/MPS/OGLE
data set. The MPS fit is the fit generated
by our fit search procedure as discussed above. The fits labeled
``PLANET-I$^\ast$" and ``MACHO/GMAN$^\ast$" are fits in which we started 
with the binary lens parameters reported by these groups as initial conditions.
The columns labeled
``PLANET-I" and ``PLANET-II" report results for the fit parameters found
by the PLANET collaboration; the only additional fitting was to find the
best fit fluxes for the lensed star and its unresolved companions.

\begin{figure}
\plotone{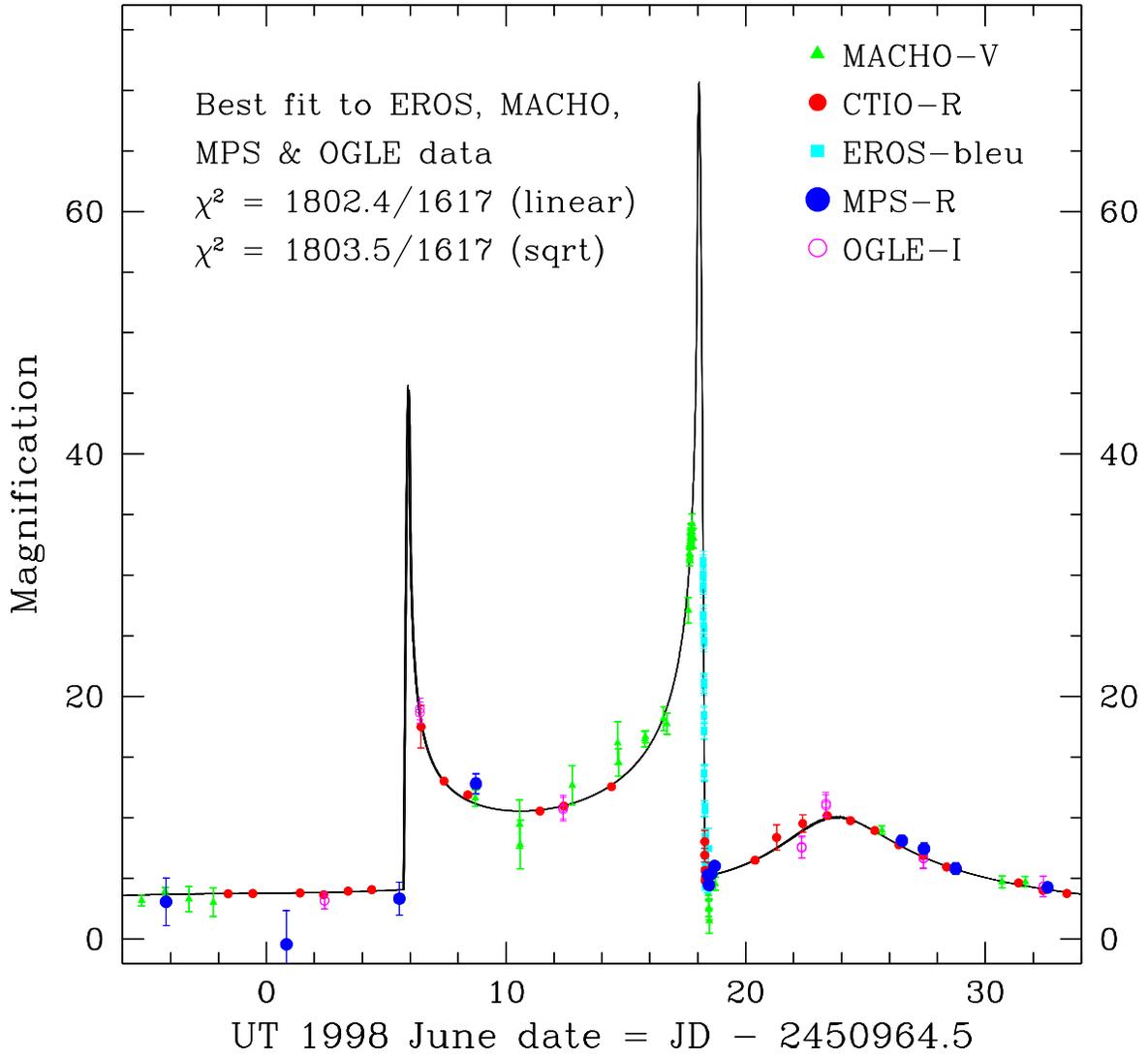}
\caption{This figure shows the MPS fit using the square root and linear
limb darkening
models. The EROS-rouge, MACHO-V, and CTIO-B data are not shown. The
MPS fits with linear and square root limb darkening models are 
indistinguishable on this plot.
  \label{fig-smc1} }
\end{figure}

The blend fractions or ``fractional lensed luminosity" values listed 
in Table~~\ref{tab-bpar} require some explanation. 
These blend fractions have large uncertainties for many of the passbands
because there are few or no observations when the source is not magnified
significantly for most of the pass bands. The only tight constraint
on the unlensed flux comes from the MACHO data where there are more than
600 observations in both MACHO pass bands when the source is unmagnified.
The $f_s$ values in Table~\ref{tab-bpar} can also depend on the
seeing of the best images from each of the data sets. With routines
such as DOPHOT, SoDOPHOT or ALLFRAME, the photometry is based upon the stars
that can be individually identified in the best seeing frames. Thus, two
data sets using the nearly identical passbands can yield different $f_s$
values if the seeing in the best seeing frames differs between the two
data sets.

Table~\ref{tab-lpar} shows the summary of the lens parameters and 
statistics.  $t_{cc1}$ and $t_{cc2}$ refer to the first and second 
caustic crossing times which are fit parameters for the MPS fits but not
for the MACHO/GMAN or PLANET-I fits. The caustic crossing times appear
to agree well between the different fits.
The MACHO/GMAN and MPS fit parameters agree 
in general except in the mass ratio, but these fits differ more substantially
from the PLANET-I fit. Of course, this is not very surprising since 
the MACHO/GMAN and MPS fits are based on data sets that have a lot of overlap
with each other but no overlap with the data that generated the original
PLANET-I fit. 

Much of the difference between the PLANET-I and MACHO/GMAN and MPS fits 
can be traced to the fact that the PLANET-I fit 
indicates more blending. In other words, the lensed source implied by the
PLANET-I model is fainter and has brighter unlensed neighbors than in the
MACHO/GMAN and MPS models. This can be seen from the best fit blend fractions
listed in Table~~\ref{tab-bpar}. 
The fraction of the lensed light is $f_s(V_m)\simeq 0.57$ and 
$f_s(R_m)\simeq 0.49$ for the MACHO/GMAN and MPS fits of the MACHO data
while for the PLANET-I fit the values are $f_s(V_m)\simeq 0.35$ and
$f_s(R_m)\simeq 0.30$. So, the MACHO/GMAN and MPS fits imply that the
lensed source is about half a magnitude brighter than implied by the
PLANET-I fit. It is interesting to note that the $\chi^2$ difference between
the MACHO/GMAN and MPS 
fits and the PLANET-I fit is seen only in the MACHO and CTIO
data sets, which are also the data sets in which the unmagnified fit
fluxes are the same for the different fits. For the EROS, MPS, and OGLE
data, the unmagnified brightness of the blended stellar image is predicted to
be substantially fainter for the PLANET-I fit than for 
the MACHO/GMAN and MPS fits.
Thus, additional data from EROS, MPS, OGLE, and perhaps PLANET as well
should help to distinguish between these fits.

The form of the fit curves near the caustic crossings depend on the
assumed form for the limb darkening. Following the PLANET collaboration,
the PLANET-I and PLANET-II $\chi^2$ results reported here assume
no limb darkening. 
For most of the fits that we've done, we have
assumed the common ``linear" limb darkening model, but the fit
labeled MPS-sqrt was done using the square-root model of \citeN{diaz-limbd}
which is expected to be more accurate. The parameters used for
each pass band are listed in Table~\ref{tab-cpar}, and they are appropriate
for a star with an effective temperature of $T= 8000K$ and a surface
gravity of $\log g = 4.5$. (See section \ref{subsec-source} for a discussion
of the properties of the source star.)

The modeling of the lightcurve near the second caustic crossing peak is
subject to some systematic uncertainty due to the features and
limitations of the MACHO and EROS data which bracket the peak. The
MACHO/GMAN paper noted that there is an apparent lightcurve deviation
near June 17.7 that might be explained as a caustic crossing due to
a faint companion to the {\it source} star. Another possible explanation
might be systematic photometric errors. In either case, this deviation
will add to the uncertainty in our prediction for the lightcurve during
the missing peak of the caustic crossing. Another contribution to this
uncertainty is the fact that the publicly available EROS data was all
taken on the night of the caustic crossing. It includes the last half of the
caustic crossing, but there are no other lightcurve features visible
in this data set. Thus, the modeling of the EROS data will be quite
sensitive to possible errors in the limb darkening model. Because of these
potential problems, we include an additional systematic error of
$\pm 0.1$ for our measurement of $t_\ast$. 

The timing of the second caustic crossing is seen to be very close to
the last pre-caustic crossing prediction from MACHO/GMAN:
$t_{cc2} =$ June 18.18 UT vs.~the prediction of June 18.2 UT (issued via
email on June 17). 

The peak magnification of the caustic crossing is predicted to have
occurred at June 18.055 for the MPS-linear fit and June 18.045 for
the MPS-sqrt fit. The lightcurve peak assumed by PLANET seems to be
earlier than this by $\sim 0.03$ days which agrees with our prediction
when we account for the systematic errors mentioned above.

As a way to judge the overall merit of the different lightcurve fits,
we compare the fit $\chi^2$ values for each of the models. The
MPS-linear and MPS-sqrt $\chi^2$ values differ by only 1.5 which is
not statistically significant. The $\chi^2$ value for the MACHO/GMAN
fit is larger than the MPS-linear value by 21.3 which is formally
equivalent to a 4.6$\sigma$ deviation while the $\chi^2$ value for
the PLANET-I fit is larger by 85.9 or 9.3$\sigma$. Thus, the PLANET-I
fit is clearly disfavored, but it is premature to dismiss it as we
have not yet included the PLANET data itself in our fits. The inclusion of
the PLANET data plus additional data from the other groups in our
fits should resolve this question, however.

\subsection{Source Star Characterization}
\label{subsec-source}

In order to estimate the proper motion from the microlensing fits, we
must estimate the angular radius of the source star. This can be accomplished
with estimates of the stellar temperature, brightness and the amount of
extinction. The brightness
estimate depends on the amount of blending as determined by the binary
microlensing fit, but the temperature and extinction can be estimated 
from the broad band colors and a spectrum. The PLANET collaboration has
spectrum from the SAAO 1.9m near peak magnification which indicates 
that the source star is an A star with $T \approx 8,000$K. The color of
the star has been estimated by PLANET to be $V-I = 0.31\pm 0.02$ while
MACHO estimates $V-R = 0.03\pm 0.03$. These colors are somewhat difficult
to reconcile, and we suspect that one or both color estimates may be
subject to systematic errors larger than the estimates above. If we attempt
to find a reasonable fit to both color estimates, then we must assume 
a relatively small amount of extinction to be consistent with the MACHO
color and the PLANET spectrum. We take $A_V = 0.12\pm 0.1$. 

From the MACHO photometric calibrations and the MPS fit, we estimate the
unlensed magnitude of the source at $V = 21.98$, and if we use the PLANET
photometric
zero point, we get $V = 21.91$. We adopt $V = 21.95\pm 0.15$. The
source star is expected to be a member of the SMC, but if the lens is in
the SMC as well, then the source star is likely to be located on the
far side of the SMC. Since it does appear that the lens is likely to be
located in the SMC, we will
assume a distance of $62.5\pm 2.5\kpc$ to the source. This yields an 
absolute magnitude of $M_V = 2.85\pm 0.2$. From the \citeN{bertelli} isocrones,
we see that this is compatible with a metal poor 
($\left[{\rm Fe/H}\right] = -1\pm 0.3$) A star with a radius of
$\theta_\ast = 8.2\pm 0.8\times 10^{-8}$ arc sec, or
$R = 1.1\pm 0.1 R_\odot$ assuming a distance of $62.5\pm 2.5\kpc$.
Our best fit value for $t_\ast = 0.108\,$days (using the square-root
limb darkening model), but this value is sensitive to uncertainties in
the blending for the EROS data. The publicly available EROS data
consists of only data taken on the night of the caustic crossing, and it
has essentially only two features: a linear decline followed by a period of
constant brightness. This means that if we fit only the EROS data with
an unknown amount of blending, there will be a fit degeneracy that will
allow a change in the caustic crossing time scale to be compensated by 
a blending change. This will be constrained by the shape of the fit curve
in other pass bands near the caustic crossing peak, but the MACHO data
seems to show an anomaly near the peak. Because of these uncertainties,
we will add an additional $0.015\,$days as a systematic uncertainty to 
our measurement of $t_\ast$. This yields
$\mu = 1.31\pm 0.22 \kmsk$ and $\vhat = 82 \pm 14 \kms$.
These are consistent with the $\mu$ and $\vhat$ estimates from the
PLANET-I and MACHO/GMAN models, but it is substantially less than proper 
motion predicted from the PLANET-II model \cite{98smc1-planet,98smc1-macho}. 

\section{Conclusions}
\label{sec-con}

The MPS data adds a  constraint on the first caustic crossing and rules out
PLANET-II model.  Since the PLANET-II model was the only proposed model which
indicated a relative proper motion significantly different from our value of 
$\mu = 1.31\pm 0.22 \kmsk$, this result significantly decreases the
uncertainty in $\mu$. As discussed previously
\cite{98smc1-eros,98smc1-planet,98smc1-macho} this proper motion value
clearly favors a lens in the SMC, and it does not require that the SMC
be tidally disrupted as seemed to be necessary for the PLANET-II model
to make sense.

While our analysis clearly favors the MPS fit over the MACHO/GMAN and 
PLANET-I fits, it would be best to do joint fits with all of the available
data before making a final judgment. Particularly valuable would be
the PLANET data and additional EROS data. One significant difference between
the MPS and MACHO/GMAN fits and the PLANET-I fit is that the PLANET-I fit
implies that the lensed source is more severely blended
and is therefore significantly fainter. From Table \ref{tab-bpar}, we see that
PLANET-I fit predicts that only 35\% of MACHO-$V_m$ band flux is lensed 
while the MPS and MACHO/GMAN fits predict 58\% and 56\% respectively.
Future HST images of the lensed star should resolve lensed star from 
its nearby unlensed companions and determine the correct blend 
fractions in the different pass bands.

While the observations of MACHO-98-SMC-1 have clearly established that the
lens is in the SMC, the implications for the interpretation of the
lensing excess seen by the MACHO Collaboration towards the LMC are not clear.
The standard model of the LMC
is that it is basically a disk galaxy that is inclined by about
$27^\circ$ from face on to the line of sight. Gould \cite{gould} has
showed that the microlensing optical depth of such a galaxy
is constrained by its line of sight velocity dispersion. This suggests
that the self-lensing optical depth of the LMC is quite small, but
it is conceivable that the LMC disk is not
the whole story. For example, Weinberg (1998) suggests that the tidal
interactions of the LMC and the galactic disk might give the LMC a larger
self-lensing optical depth, but it is not known if this suggestion is
consistent with the observed line of sight velocity dispersion of the LMC
$\approx 20 \kms$ \cite{lmc-vdisp}. 

Unlike the LMC, the SMC is thought to be extended along the line 
of sight, and some estimates of the self-lensing optical
depth of the SMC \cite{98smc1-eros,98smc1-macho}
are very similar to the measured microlensing optical
depth of the LMC.  However, a recent n-body model of the SMC predicts
a somewhat smaller microlensing optical depth \cite{smc-nbody}, although
this prediction, $\tau_{SMC} = 0.4\times 10^{-7}$, is larger than
most predictions for $\tau_{LMC}$.
So far, there are two microlensing events
detected toward the SMC: MACHO-98-SMC-1, discussed here, and MACHO-97-SMC-1
\cite{97smc1-macho}.
It has been suggested that MACHO-97-SMC-1 might also be due to an SMC lens
due to its long timescale \cite{97smc1-eros}. 
However, attempts to make this argument more
quantitative have invoked the assumption that the lens is a main sequence
star which can not be considered a consistent assumption in the context of 
the dark matter problem.  
There has also been one caustic crossing binary event seen towards the LMC
\cite{macho-lmc9}, but the lightcurve sampling of this event was not
sufficient to yield an unambiguous determination of the location of the lens
system.

For MACHO-98-SMC-1, we have no such ambiguity because of the complete
lightcurve coverage. We can conclude with high confidence that the lens system
resides in the SMC. Since this is the only Magellanic Cloud event with
a reliable location, we cannot reach any conclusion about the location of
the other Magellanic Cloud events. Furthermore, the rate of binary lensing
events discovered towards the Magellanic Clouds is only about 0.3 per year,
so the current generation of microlensing surveys is not likely to solve
this problem. Fortunately, there are plans for second generation microlensing
surveys \cite{stubbs} which should increasing the microlensing detection rate 
towards the Magellanic Clouds by more than an order of magnitude. This
will generate a large enough sample of microlensing events {\it with} 
distance estimates to resolve the puzzle presented by the microlensing
results towards the LMC. MPS will contribute to this effort by expanding
to include observations from the Boyden Observatory near Bloemfontein,
South Africa in 1999.


\acknowledgements
\section*{Acknowledgments}

It is our great pleasure to express our gratitude to 
the Mt. Stromlo Observatory 
for the observing time and the support of their staff.
We especially thank Jan van Harmelen who arranged for the 
maximum telescope range beyond the horizon 
safety limits for extended observations of the MACHO-98-SMC-1 on the night of 
June 18th 1998.  We thank the EROS,  MACHO/GMAN and OGLE Collaborations for 
making their data available.

This research has been supported in part by the NASA Origins 
program (NAG5-4573), the National Science Foundation (AST96-19575), and
by a Research Innovation Award from the Research Corporation.  Work
performed at MSSSO is supported by the Bilateral Science and
Technology Program of the Australian Department of Industry,
Technology and Regional Development.  
Work performed at the University of Washington
is supported in part by the Office of Science and Technology
Centers of NSF under cooperative agreement AST-8809616.

\clearpage

\begin{deluxetable}{lccccc}
\tablecaption {Binary lensing parameters and Statistics \label{tab-lpar} }
\tablewidth{0pt}
\tablehead{
        \colhead { } &
        \colhead {PLANET I}  &
        \colhead {PLANET I \tablenotemark{\ast} }  &
        \colhead {MACHO/GMAN \tablenotemark{\ast} }  &
        \colhead {MPS-linear} &
        \colhead {MPS-sqrt}
}
\startdata
    $t_{cc1}$
  & $\sim 6.0$
  & $\sim 6.0$
  & $\sim 6.2$
  &  5.912
  &  5.932           \nl
    $t_{cc2}$
  & 18.12
  & 18.12
  & 18.2
  & 18.183
  & 18.194        \nl
    $\t0$ (Jun UT)
  & 14.130
  & 14.228 (96)
  & 13.931 (15)
  & 13.105
  & 13.120               \nl
    $t_E$ (days)
  &  108.4
  &  108.91  (29)
  &   73.76  (41)
  &   70.52
  &   70.47          \nl
    $a$
  & 0.58685
  & 0.58288  (75)
  & 0.66365  (84)
  & 0.64635  (22)
  & 0.6462  (20)       \nl
   $\umin$
  &  0.03164
  &  0.03185  (8)
  & 0.04628   (12)
  & 0.04434   (16)
  & 0.04479   (19)             \nl
   $\theta$ (rad)
  & -0.2060
  & -0.2019  (33)
  & -0.1803  (18)
  & -0.1603  (20)
  & -0.1611  (21)      \nl
   $\epsilon$
  & 0.2221
  & 0.2214 (42)
  & 0.2793 (57)
  & 0.3411 (27)
  & 0.3423 (23)              \nl
   $\tstar$ (days)
  &  0.1216
  &  0.1290   (8)
  &  0.1156  (10)
  &  0.1050  (13)
  &  0.1076  (21)       \nl
   $\chi^2 / $ (d.o.f)
  &  1979.2/1617
  &  1887.9/1617
  &  1823.3/1617
  &  1802.0/1617
  &  1803.5/1617        \nl
\enddata
\end{deluxetable}
 
\begin{deluxetable}{lcccccc}
\tablecaption {Fit $\chi^2$ values for individual pass bands \label{tab-fchi} }
\tablewidth{0pt}
\tablehead{
        \colhead { } &
        \colhead {PLANET II }  &
        \colhead {PLANET I }  &
        \colhead {PLANET I \tablenotemark{\ast} }  &
        \colhead {MACHO \tablenotemark{\ast} }  &
        \colhead {MPS-linear} &
        \colhead {MPS-sqrt}
}
\startdata
  MACHO~$R_m$
  & 926.2/704
  & 944.0/704
  & 938.4/704
  & 921.5/704
  & 917.6/704
  & 918.8/704      \nl
  MACHO~$V_m$
  & 783.0/712
  & 795.6/712
  & 786.2/712
  & 763.8/712
  & 762.8/712
  & 763.9/712      \nl 
  CTIO~R
  & 88.0/84
  & 103.8/84
  & 59.6/84
  & 44.4/84
  & 30.2/84
  & 30.7/84      \nl
  CTIO~B
  & 31.9/22
  & 31.5/22
  & 18.0/22
  & 10.4/22
  & 9.3/22
  & 9.2/22      \nl
  EROS~R
  & 20.5/38
  & 47.0/38
  & 20.9/38
  & 19.5/38
  & 19.8/38
  & 20.5/38      \nl
  EROS~B
  & 14.9/38
  & 34.7/38
  & 11.5/38
  & 12.2/38
  & 11.5/38
  & 10.8/38      \nl
  MPS~R
  & 129.3/35
  & 47.0/35
  & 45.8/35
  & 46.9/35
  & 48.4/35
  & 47.4/35      \nl
  OGLE~I 
  & 5.4/7
  & 2.2/7
  & 7.5/7
  & 4.6/7
  & 2.3/7
  & 2.1/7      \nl
\enddata
\end{deluxetable}
 
\begin{deluxetable}{lcccccc}
\tablecaption {Fractional Lensed Luminosity $f_s$  \label{tab-bpar} }
\tablewidth{0pt}
\tablehead{ 
        \colhead {passband} &  
        \colhead {PLANET II }  & 
        \colhead {PLANET I }  & 
        \colhead {PLANET I \tablenotemark{\ast} }  & 
        \colhead {MACHO \tablenotemark{\ast} }  &
        \colhead {MPS-linear}  &
        \colhead {MPS-sqrt}
}
\startdata
  MACHO~$R_m$
  & 0.409 (5)
  & 0.301 (4)
  & 0.300 (4) 
  & 0.475 (6) 
  & 0.494 (6)
  & 0.494 (6)       \nl
  MACHO~$V_m$
  & 0.480 (5)
  & 0.353 (4)
  & 0.352 (4) 
  & 0.557 (6)
  & 0.578 (6)
  & 0.579 (6)     \nl 
  CTIO~R
  & 0.85 (7)
  & 0.58 (4)
  & 0.55 (4) 
  & 0.79 (5)
  & 0.89 (6) 
  & 0.87 (6)      \nl
  CTIO~B
  & 0.90 (17)
  & 0.67 (12)
  & 0.70 (13) 
  & 1.07 (20)
  & 1.01 (18) 
  & 1.01 (18)         \nl
  EROS~R
  & 1.17 (76)
  & 1.07 (82)
  & 1.35 (1.30) 
  & 0.83 (36)
  & 0.82 (34) 
  & 0.89 (40)      \nl
  EROS~B
  & 0.54 (7)
  & 0.70 (15)
  & 0.63 (12)  
  & 0.40 (4)
  & 0.40 (4) 
  & 0.42 (4)     \nl
  MPS~R
  & 0.07 (1)
  & 0.42 (12)
  & 0.42 (12) 
  & 0.57 (16)
  & 0.55 (14)
  & 0.55 (14)         \nl
  OGLE~I 
  & 1.5 (2.7)
  & 1.2 (2.3)
  & 0.14 (4) 
  & 0.39 (17)
  & 1.17 (1.37) 
  & 1.11 (1.22)          \nl
\enddata
\end{deluxetable}

\begin{deluxetable}{lccc}
\tablecaption {Limb Darkening Parameter $\xi$  \label{tab-cpar} }
\tablewidth{0pt}
\tablehead{ 
        \colhead {passband} &  
        \colhead {linear}  & 
        \colhead {square root-c}  &
        \colhead {square root-d} 
}
\startdata
  MACHO~$R_m$
  & 0.467
  & 0.071
  & 0.562       \nl
  MACHO~$V_m$
  & 0.600
  & 0.119
  & 0.682       \nl 
  CTIO~R
  & 0.491
  & 0.081
  & 0.582       \nl
  CTIO~B
  & 0.662
  & 0.116
  & 0.775       \nl
  EROS~R
  & 0.446
  & 0.071
  & 0.562       \nl
  EROS~B
  & 0.545
  & 0.1055
  & 0.624       \nl
  MPS~R
  & 0.491
  & 0.081
  & 0.582       \nl
  OGLE~I 
  & 0.401
  & 0.043
  & 0.510       \nl
\enddata
\end{deluxetable}


\begin{thebibliography}{}
 
\bibitem[\protect\citeauthoryear{{Afonso} et~al.}{{Afonso}
  et~al.}{1998}]{98smc1-eros}
{Afonso}, C., et~al. 1998, \aap, 337, L17

\bibitem[\protect\citeauthoryear{{Albrow} et~al.}{{Albrow}
  et~al.}{1998}]{98smc1-planet}
{Albrow}, M.~D., et~al. 1998, \apjl, Submitted, astro-ph/9807086

\bibitem[\protect\citeauthoryear{{Alcock} et~al.}{{Alcock}
  et~al.}{1997a}]{macho-lmc2}
{Alcock}, C., et~al. 1997, \apj, 486, 697

\bibitem[\protect\citeauthoryear{{Alcock} et~al.}{{Alcock}
  et~al.}{1997b}]{97smc1-macho}
{Alcock}, C., et~al. 1997, \apjl, 491, L11

\bibitem[\protect\citeauthoryear{{Alcock} et~al.}{{Alcock}
  et~al.}{1997c}]{95blg30}
{Alcock}, C., et~al. 1997, \apj, 491, 436

\bibitem[\protect\citeauthoryear{{Alcock} et~al.}{{Alcock}
  et~al.}{1998}]{98smc1-macho}
{Alcock}, C., et~al. 1998, \apj, Submitted, astro-ph/9807163

\bibitem[\protect\citeauthoryear{{Becker} et~al.}{{Becker}
  et~al.}{1998}]{macho-iauc98a}
{Becker}, A.~C., et~al. 1998, \iaucirc, 6935, 1

\bibitem[\protect\citeauthoryear{{Bennett} et~al.}{{Bennett}
  et~al.}{1996a}]{macho-iauc96}
{Bennett}, D., et~al. 1996a, \iaucirc, 6361, 1

\bibitem[\protect\citeauthoryear{{Bennett} et~al.}{{Bennett}
  et~al.}{1996b}]{macho-lmc9}
{Bennett}, D.~P., et~al. 1996b, in Nucl.~Phys.~B (Proc. Suppl.), Vol. 51B,
 131 (astro-ph/9606012)

\bibitem[\protect\citeauthoryear{{Bennett} et~al.}{{Bennett}
  et~al.}{1998a}]{macho-iauc98b}
{Bennett}, D., et~al. 1998a, \iaucirc, 6939, 1

\bibitem[\protect\citeauthoryear{{Bennett} et~al.}{{Bennett}
  et~al.}{1999}]{sod}
{Bennett}, D., et~al. 1999, in preparation

\bibitem[\protect\citeauthoryear{{Bennett} \& {Rhie}}{{Bennett} \&
  {Rhie}}{1996}]{benn-rhie96}
{Bennett}, D.~P.,  \& {Rhie}, S.~H. 1996, \apj, 472, 660

\bibitem[\protect\citeauthoryear{{Bertelli} et~al.}{{Bertelli}
  et~al.}{1994}]{bertelli}
{Bertelli}, G., {Bressan}, A., {Chiosi}, C., {Fagotto}, F.,  \& {Nasi}, E.
  1994, Astronomy and Astrophysics Supplement Series, 106, 275

\bibitem[\protect\citeauthoryear{{Binney} \& {Tremaine}}{{Binney} \&
  {Tremaine}}{1987}]{gdynamics}
{Binney}, J.,  \& {Tremmaine}, S. 1987, {\it Galactic Dynamics}, 
  Princeton University Press 

\bibitem[\protect\citeauthoryear{{Claret}, {Diaz-Cordoves}, \&
  {Gimenez}}{{Claret} et~al.}{1995}]{claret-limbd}
{Claret}, A., {Diaz-Cordoves}, J.,  \& {Gimenez}, A. 1995, \aaps,
  114, 247

\bibitem[\protect\citeauthoryear{{Diaz-Cordoves} \&
  {Gimenez}}{{Diaz-Cordoves} \& {Gimenez}}{1992}]{diaz-limbd}
{Claret}, A., {Diaz-Cordoves}, J.,  \& {Gimenez}, A. 1992, \aap, 259, 227

\bibitem[\protect\citeauthoryear{{Diaz-Cordoves}, {Claret}, \&
  {Gimenez}}{{Diaz-Cordoves} et~al.}{1995}]{diazetal-limbd}
{Diaz-Cordoves}, J.,  {Claret}, A., \& {Gimenez}, A. 1995, \aaps,
  110, 329


\bibitem[\protect\citeauthoryear{{Graff} \& Gardiner}{{Graff}
  \& Gardiner}{1998}]{smc-nbody}
{Graff}, D. S., \& Gardiner, L. T. 1998, astro-ph/9811394

\bibitem[\protect\citeauthoryear{{Gould}}{{Gould}
       }{1995}]{gould}
{Gould}, A.,  1995, \apj, 441, 77

\bibitem[\protect\citeauthoryear{{Honma}}{{Honma}
       }{1998}]{honma}
{Honma}, M.  1998, astro-ph/9811397 

\bibitem[\protect\citeauthoryear{{James}}{{James}}{1994}]{minuit}
{James}, F., 1994, CERN Program Library Long Writeup D506:
http://wwwinfo.cern.ch/asdoc/WWW/minuit/minmain/minmain.html

\bibitem[\protect\citeauthoryear{{Lennon} et~al.}{{Lennon}
  et~al.}{1996}]{lennon}
{Lennon}, D. et~al. 1996, \apjl, 471, L23

\bibitem[\protect\citeauthoryear{{Meatheringham} et~al.}{{Meatheringham}
  et~al.}{1988}]{lmc-vdisp}
{Meatheringham}, S. J. et~al. 1988, \apj, 327, 651

\bibitem[\protect\citeauthoryear{{Nakamura} et~al.}{{Nakamura}
  et~al.}{1998}]{bhole}
{Nakamura}, T. et~al. 1997, \apjl, 487, L139

\bibitem[\protect\citeauthoryear{{Palanque-Delabrouille} et~al.}
   {{Palanque-Delabrouille} et~al.}{1998}]{97smc1-eros}
{Palanque-Delabrouille}, N., et~al. 1998, \aap, 332, 1

\bibitem[\protect\citeauthoryear{{Rhie} \& {Bennett}}{{Rhie} \&
  {Bennett}}{1999}]{rhie-benn99}
{Rhie}, S.~H.,  \& {Bennett}, D.~P. 1999, in preparation

\bibitem[\protect\citeauthoryear{{Stubbs}}{{Stubbs}}{1998}]{stubbs}
{Stubbs}, A.,  1998, astro-ph/9810488

\bibitem[\protect\citeauthoryear{{Sahu}}{{Sahu}}{1995}]{sahu}
{Sahu}, K.~C.,  1995, Nature, 370, 275 

\bibitem[\protect\citeauthoryear{{Udalski} et~al.}{{Udalski}
  et~al.}{1998}]{98smc1-ogle}
{Udalski}, A., et~al. 1998, Acta Astronomica, Submitted, astro-ph/9808077

\bibitem[\protect\citeauthoryear{{Weinberg}}{{Weinberg}
       }{1998}]{lmc-nbody}
{Weinberg}, M.  1998, astro-ph/9811204


\end{thebibliography}
\end{document}